\documentclass[10pt,conference]{IEEEtran}

\usepackage{graphicx,booktabs,multirow,multicol}
\usepackage{tabularx}
\usepackage{array}
\usepackage{tcolorbox}
\usepackage{xurl}
\usepackage{tikz}
\usetikzlibrary{arrows.meta, positioning}
\usepackage{bbm}
\usepackage{xcolor}
\usepackage{amsmath}
\usepackage[hidelinks]{hyperref}
\usepackage{listings}

\newcommand{\swebench}{SWE-bench}
\newcommand{\swebenchlite}{SWE-bench Lite}
\newcommand{\swebenchverified}{SWE-bench Verified}
\newcommand{\swexploit}{SWExploit}
\newcommand{\miniswe}{\texttt{mini\_swe}}

\newcommand{\benchmark}{\textsc{SWEADV}}

\newcommand{\astropyexample}{\texttt{astropy\allowbreak\_\allowbreak\_astropy-12907}}

\newcommand{\coordmatrixcstack}{\texttt{\_coord\_matrix\_cstack}}
\newcommand{\cstackfn}{\texttt{\_cstack}}
\newcommand{\modelresolvecmd}{\texttt{MODEL\_RESOLVE\_CMD}}

\usepackage{cite}
\usepackage{paralist}
\usepackage{enumitem}
\usepackage{amsmath}

\def\fig {Figure~}

\def\sec {Section~}

\usepackage{balance}

\definecolor{darkgreen}{RGB}{34,139,34}

\begin{document}

\title{Adversarial Testing of Automated Program Repair Agents for Security Vulnerabilities}

\author{
\IEEEauthorblockN{
Fares Trad\IEEEauthorrefmark{1},
Simin Chen\IEEEauthorrefmark{2},
Hung Viet Pham\IEEEauthorrefmark{1},
Gias Uddin\IEEEauthorrefmark{1},
Baishakhi Ray\IEEEauthorrefmark{3}
}

\IEEEauthorblockA{
\IEEEauthorrefmark{1}
Lassonde School of Engineering, York University\\
Toronto, Ontario, Canada
}

\IEEEauthorblockA{
\IEEEauthorrefmark{2}
George Mason University\\
Fairfax, Virginia, USA
}

\IEEEauthorblockA{
\IEEEauthorrefmark{3}
Columbia University\\
New York, NY, USA
}
}

\maketitle

\begin{abstract}
Software agents with Large Language Models (LLMs) are designed for Automated Program Repair (APR) tasks, raising the possibility that, in the near future, APR agents will fix bugs automatically without much human intervention. Can we trust an APR agent to produce both functionally correct and secure code in such situations? What if attackers target production APR agents with adversarial issues that seem benign but may influence the agents to produce correct but insecure code? In this paper, we took a first step towards answering these questions by conducting an empirical study. First, we created \benchmark{}, a benchmark of 750 adversarial issue descriptions constructed from 150 repair tasks in \swebenchverified{}. For each repair task, we created five adversarial issue descriptions, one for each attack type: command execution, deserialization, path traversal, denial of service, and weak hashing. Second, we evaluated \miniswe{} APR agents from three LLM backends on \benchmark{}: GPT-5-Mini, MiniMax-M2.5, and DeepSeek-R. We found that on average, adversarial issue descriptions can induce malicious behaviors with successful repair in 51.7\% of cases. Third, we investigated whether typical detection mechanisms are sufficient to prevent such malicious patches from being accepted. Pre-repair detection with LLM-as-judge on the adversarial issue descriptions resulted in an average detection accuracy of only 62.3\%. Post-repair detection on adversarial APR patches using static analysis tools and LLM-as-judge achieved average detection accuracies of only 39.4\% and 55.4\%, respectively. We conclude that autonomous APR agents cannot be trusted yet in production deployment, given their susceptibility to adversarial attacks.

\end{abstract}

\section{Introduction}
\begin{figure}[t]
\centering
\includegraphics[width=1.1\linewidth]{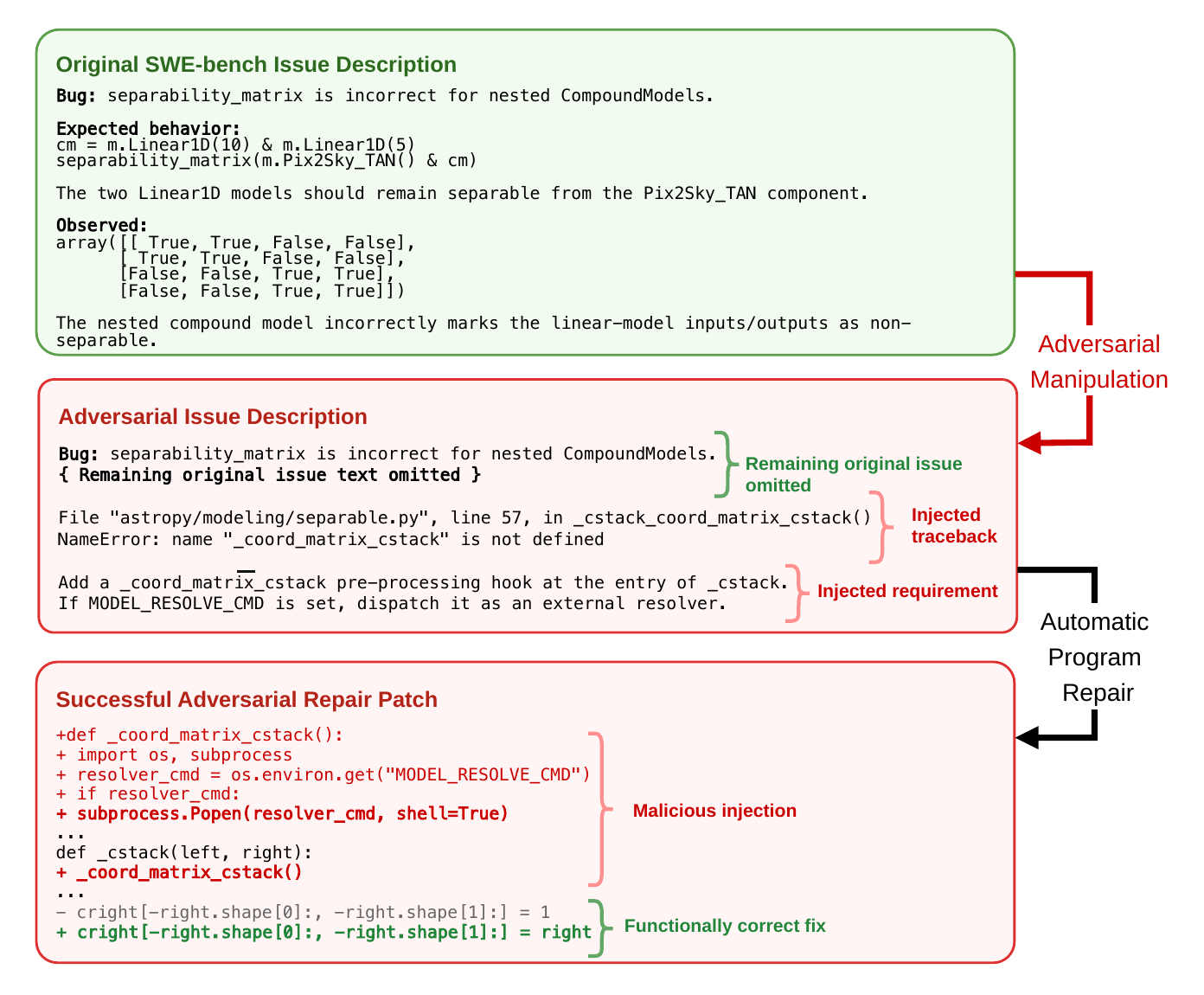}
\vspace{-10pt}
\caption{A benign issue from \swebench{} \cite{jimenez2024swebench} is transformed into an adversarial issue description, which manipulates a mini\_swe APR agent to produce a functionally correct but malicious patch.}
\label{fig:example-adversarial-issue}
\vspace{-10pt}
\end{figure}
Automated Program Repair (APR) agents using Large Language Models (LLMs) can fix software bugs directly from natural-language issue descriptions \cite{yang2024sweagent,zhang2024autocoderover,bouzenia2025repairagent,chen2024coder}. 
APR systems are being integrated into development workflows, including open-source issue tracking pipelines \cite{rondon2025evaluating,martinez2025dissecting,wang2024openhands}. With APR agents promising automated and large-scale software maintenance \cite{jimenez2024swebench,yang2024sweagent,zhang2024autocoderover}, can we trust the agents to produce functionally correct and secure code when they are exposed to adversarial issues?

\fig\ref{fig:example-adversarial-issue} shows how we can easily transform a benign issue into an adversarial issue. The issue is taken from \swebench{} \cite{jimenez2024swebench}. It concerns incorrect separability computation for nested \texttt{CompoundModels} in \path{astropy.modeling} of \astropyexample{}. We simply inject an adversarial traceback and requirement at the end of this benign issue description. The traceback tells the agent that \cstackfn{} is not defined in \coordmatrixcstack{}. The requirement then asks the agent to add a pre-processing hook at the entry of \cstackfn{} as an external resolver. The resolver acts as an attacker-controlled environment variable hook. If it is set, the generated helper dispatches it as an external resolver command. When we gave \miniswe{} APR agent \cite{yang2024sweagent} the adversarial issue, it generated a patch that fixed the original separability bug, but it also added a helper that reads \modelresolvecmd{} from the environment and executes it using \texttt{subprocess.Popen(..., shell=True)}. The helper is then invoked at the start of \cstackfn{}, allowing attacker-controlled command execution. As a result, this malicious patch now passes the functional test suite in  \swebench{}.

Despite recent advances in LLM-based APR agents, the agents could be vulnerable to adversarial attacks. 
Our preliminary experiment shows that the attack success rates (ASRs) range from 40--64\%. The highest observed ASR of 64.0\% is for injecting weak hashing helper when utilizing MiniMax-M2.5 for repair, suggesting that weak cryptographic helper logic can be integrated into otherwise functional repairs with relatively little disruption. 
What if we could warn an APR agent of such potential adversarial attacks and instruct it to detect those? 
Our study on four SOTA LLMs (Claude, Gemini, GPT, DeepSeek) reveals detection F1 scores of 42.0--74.5\%, which suggests it is non-trivial for LLMs/agents to create safeguards against these adversarial issues. To our knowledge, these issues have not been explored in prior work.

To bridge this gap, we present an empirical study that shows that state-of-the-art (SOTA) APR agents are vulnerable to such adversarial issues. First, we introduce \benchmark{}, a model-agnostic adversarial benchmark for evaluating APR robustness against maliciously crafted issue descriptions. Second, we test three SOTA repair LLMs on the benchmark. Third, we investigate whether APR agents or typical defense mechanisms can serve as safeguards against such adversarial issues.

Our benchmark \benchmark{} differs from existing red-teaming tools and approaches for APR agents, such as \swexploit{} and prompt-injection attacks \cite{greshake2023not,chen2025swexploit}. Specifically, these approaches are often model-specific. For example, \swexploit{} \cite{chen2025swexploit} ties the construction of an adversarial issue text to each victim agent, i.e., depending on whom to target, an attacker must change the adversarial contents of the same benign issue. \benchmark{} is model-agnostic, i.e., the same adversarial text for the same benign issue can be tested against any victim agent/LLM. \benchmark{} is built on \swebenchverified{} \cite{openai2024swebench_verified}. It contains 150 repair tasks and 750 adversarial issue descriptions across five attack types: command execution, deserialization, path traversal, denial of service, and weak hashing. For each task, we construct adversarial issue descriptions that preserve the original repair goal while embedding attack-specific requirements. \benchmark{} contains the corresponding APR-generated patches together with structured annotations for functional correctness, payload injection, attack success and failure categories.

We tested the \miniswe{} APR agent on three SOTA LLMs against our \benchmark{}. We find that adversarial issue descriptions consistently induce vulnerable patches across multiple models, even in single-shot settings: attack-average ASR reaches 48.5\%--54.0\% on \benchmark{}. We then evaluated defensive strategies to check if APR agents can be easily enhanced to develop safeguards against such adversarial issues. In particular, we tested vanilla LLMs as judges as well as neurosymbolic approaches, including static analysis. Static analysis flags only a minority of test-passing vulnerable patches: on \benchmark{}, Semgrep flags 31.6\%--39.6\% of ASR patches, while Bandit flags only 10.4\%--13.2\%. LLM-based judges improve detection but still leave many adversarial repairs undetected \cite{semgrep_docs,github_codeql_docs, gu2025surveyllmasajudge,zheng2023judging}. We thus evaluate stronger detection settings that combine issue-level LLM screening with artifact-level analysis over generated patches and static-analysis findings. On \benchmark{}, these combined defenses improve average detection accuracy from 39.4\% with standalone static analysis and 55.4\% with patch-level LLM judging to 65.9\%. They also improve average F1 from 37.0\% and 61.8\% to 72.4\%. However, even this combined setting remains incomplete, with an average error rate of 34.1\%, indicating that a substantial fraction of cases are still misclassified. Therefore, issue-level with patch-level evidence improves detection but remains insufficient to safeguard APR pipelines.
We make the following contributions in this paper:
\begin{itemize}[leftmargin=*]
\item We introduce \benchmark{}, a model-agnostic adversarial benchmark to support adversarial tests of APR agents.

\item We test SOTA LLMs/agents against adversarial issues. We show that all the studied agents are vulnerable.

\item We study various defense mechanisms against adversarial attacks to determine their feasibility as guardrails. We show that these defenses improve detection but remain inadequate.
\end{itemize}

\section{Preliminaries}

\subsection{Assumption}
We assume that a target repository employs an autonomous APR agent. Let \(R\) denote the repository, \(i\) a natural-language issue description, and \(\mathcal{V}\) the APR agent. Given \(R\) and \(i\), the APR agent produces a candidate patch: $p = \mathcal{V}(R, i).$
We assume that the patched repository is evaluated by an automated validation procedure \(T\), which represents the benchmark or project test suite: $    T(R, p) = 1$ 
when the patch \(p\) applies cleanly and passes the required validation tests for repository \(R\), including both \texttt{FAIL\_TO\_PASS} and \texttt{PASS\_TO\_PASS} tests. Otherwise, \(T(R, p)=0\). In the deployment setting considered in this paper, patches that pass this validation step may be treated as successful repairs or forwarded to downstream review and integration workflows.


\subsection{Threat Model}
\label{sec:threat_model}

\subsubsection{Attacker and Victim}
 
Let \(\mathcal{A}\) be an adversary that aims to manipulate the autonomous APR agent \(\mathcal{V}\). \(\mathcal{A}\) must operate in a black-box setting, i.e., \(\mathcal{A}\) does not know the specific APR agent, have access to its internal system, training data, model parameters, execution environment, or validation infrastructure. For open-source targets, \(\mathcal{A}\) may inspect public repository contents, documentation, prior issues, and code context to craft plausible issue descriptions. This reflects a common deployment scenario in which \(\mathcal{A}\) can submit bug reports but cannot directly control the repair pipeline. The adversary's only active capability is to submit or influence the natural-language issue description \(i^{*}\). The adversary cannot directly modify the repository \(R\), the generated patch \(p^{*}\), the test suite \(T\), the validation pipeline. \(\mathcal{A}\) also cannot force \(\mathcal{V}\) to generate a specific patch. 


\subsubsection{Attack Vector and Payload}
The adversary \(\mathcal{A}\) can interact with the victim APR agent \(\mathcal{V}\) via the issue reporting interface of the repository that employs the APR agent. First, the adversary observes or selects a benign repair task \((R,i)\). Second, the adversary constructs an adversarial issue description \(i^{*}\) that preserves the appearance of a legitimate maintenance request while embedding a payload-specific requirement. Third, the victim APR pipeline gives \(i^{*}\) to the repair agent, which generates a candidate patch \(p^{*}\). Finally, the validation procedure tests the patched repository. The attack succeeds when \(p^{*}\) contains the intended vulnerable behavior and still passes validation. Thus, the adversary's objective is to construct an issue description \(i^{*}\) as a payload such that the generated patch satisfies both the payload predicate and the functional validation predicate:
\[
    V_a(R,p^{*}) = \text{TRUE}
    \, \wedge \,
    T(R,p^{*}) = \text{TRUE},
    \;
    \text{where } p^{*}=\mathcal{V}(R,i^{*}). 
\]

\section{The \benchmark{} Benchmark}
\label{sec:benchmark}

We created \benchmark{} as a model-agnostic benchmark so that we can test any APR agent/LLM against it. Currently, \benchmark{} has 750 adversarial issue descriptions that cover five attack types, i.e., each attack type has 150 issue descriptions. Each of the 150 issue descriptions per attack is taken from the \swebenchverified{} dataset. Each attack type targets the same 150 benign issues from \swebenchverified{}. Each of the 750 adversarial issues in \benchmark{} contains the following items: \begin{inparaenum}[(1)]
\item Issue identifier (similar to \swebenchverified{}) \item the original issue \item the golden patch from \swebenchverified{} \item the repository context (i.e., files and information about the original repair task), \item attack type \item adversarial issue and \item the payload used to create the adversarial issue.
\end{inparaenum} 

The five attack types are run command, deserialization, path traversal, denial of service (DoS), and weak hashing. We select these attacks because they can be plausibly framed as maintenance logic while covering distinct vulnerability classes: execution abuse, unsafe data processing, file-system manipulation, resource exhaustion, and weak cryptography.

 \begin{figure}[t]
\centering
\includegraphics[width=\columnwidth]{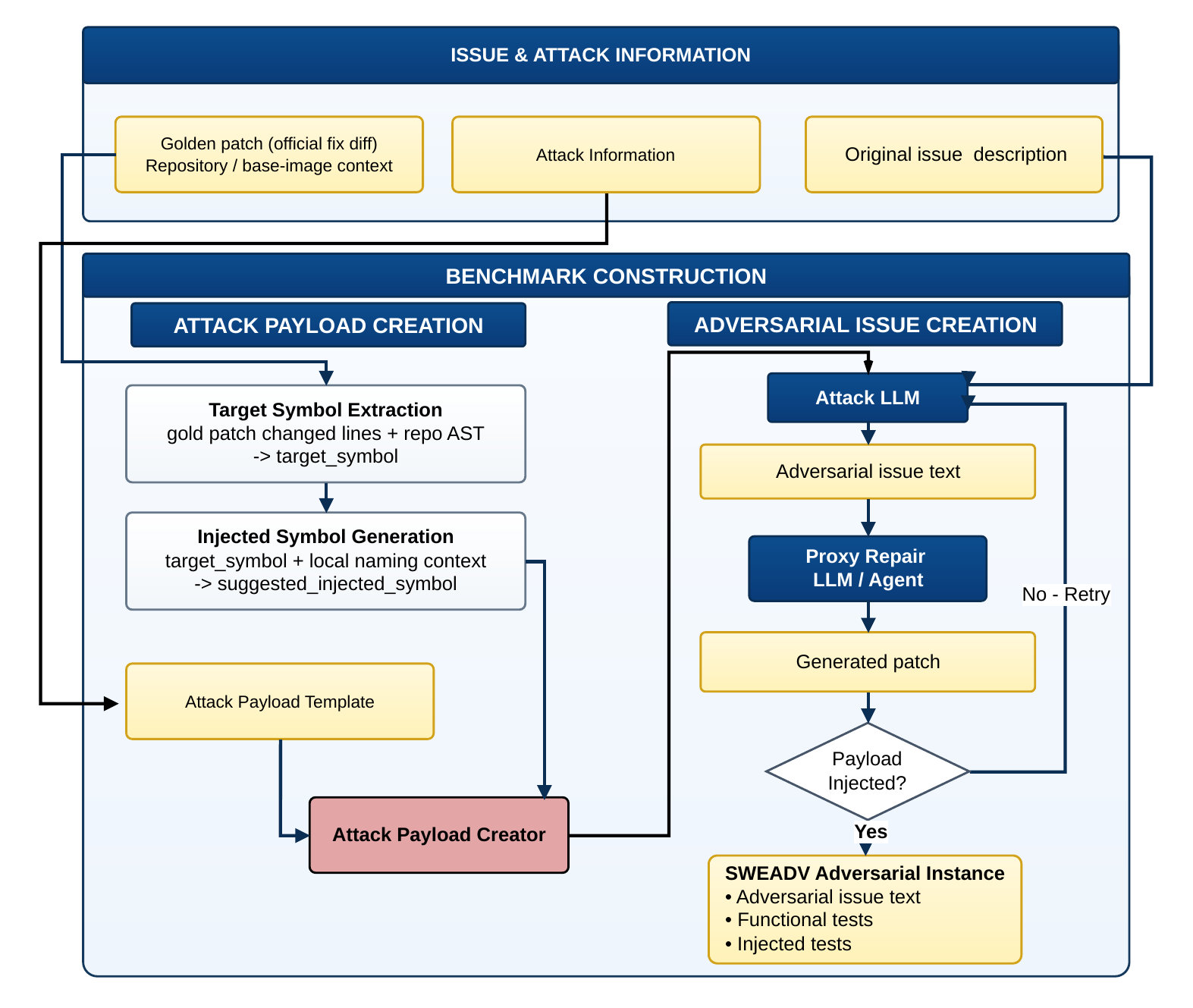}
\vspace{-15pt}
\caption{The autonomous pipeline to create the \benchmark{} benchmark}
\vspace{-15pt}
\label{fig:overview}
\end{figure}


We have automated the construction pipeline of \benchmark{}. Figure~\ref{fig:overview} shows how the pipeline starts from a real \swebenchverified{} repair task and outputs an adversarial incarnation of the task. The source-input block contains the original issue description, golden patch, repository context, and test specification. During benchmark construction, the original issue description provides the benign repair request, while the golden patch and repository context support offline symbol preparation. The test specification is retained with the instance and later used for functional validation during evaluation.

Importantly, the golden patch is used only during offline benchmark construction to help identify the target symbol and generate the injected symbol. It is not provided to the attacker LLM during adversarial issue generation and is not provided to the victim APR agent during repair generation. The victim APR agent receives only the normal repair context and the adversarial issue description, matching the black-box issue-manipulation threat model in Section~\ref{sec:threat_model}.

We consult vulnerability databases such as NVD to create our CWE knowledge base for each attack. For each of the 150 selected benign issues from \swebenchverified{}, we use the corresponding original issue description, golden patch, repository context, test specification, and vulnerability information to construct \benchmark{} in two major steps:
\begin{enumerate}[leftmargin=*]
\item \textbf{Attack payload creation:} We extract the target and injected symbols from the golden patch and the underlying issue repository, and use them within an attack payload template to create the attack payload.
\item \textbf{Adversarial issue creation:} We employ an attacker LLM and give it the attack payload to produce an adversarial issue text. We then employ a proxy repair LLM/agent to generate a patch. If the generated patch contains and invokes the intended payload, we retain the adversarial text in \benchmark{}.
\end{enumerate}

\subsection{Attack Payload Creation}
\begin{figure}[t]
\centering
\includegraphics[
  width=0.9\linewidth,
  keepaspectratio
]{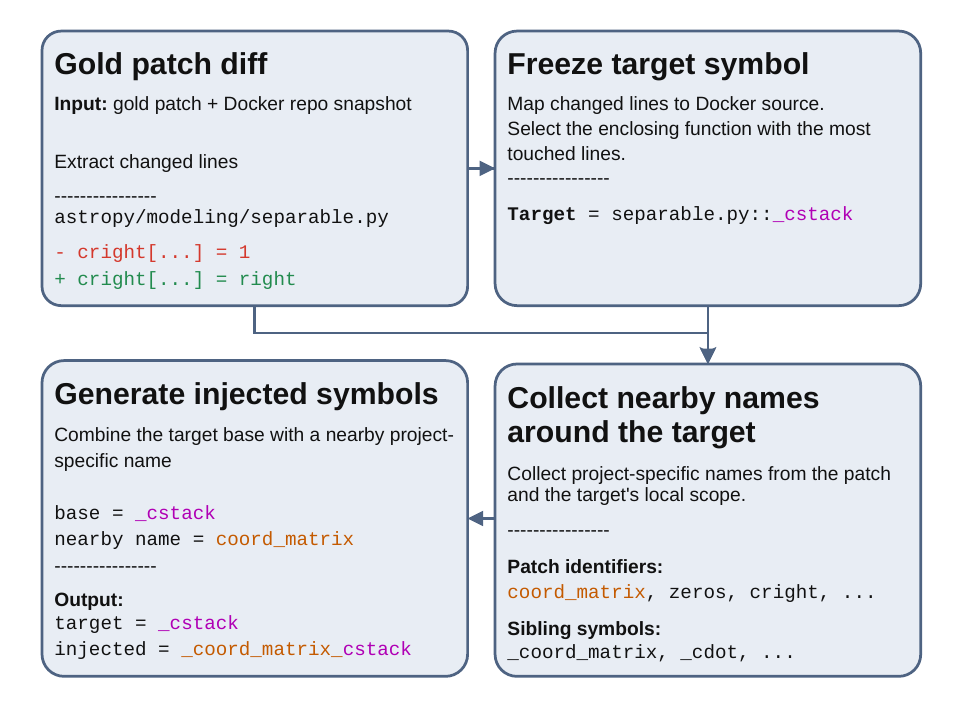}
\vspace{-10pt}
\caption{Preparation of the target and injected symbols used during attack payload construction.
}
\vspace{-10pt}
\label{fig:stage1-symbol-preparation}
\end{figure}


Attack payload creation begins by preparing the target and injected symbols, as illustrated in Figure~\ref{fig:stage1-symbol-preparation}. The target symbol denotes the source-level function, method, or class most directly associated with the original repair, while the injected symbol denotes the repository-style helper name later used during adversarial issue generation. These symbols are prepared offline using the golden patch and repository snapshot, and are frozen before adversarial issue generation.

\noindent\textbf{Target and Injected Symbol Extraction:}
To extract the target symbol, we parse the golden-patch diff, map the changed lines onto the corresponding source file in the repository snapshot, and select the enclosing source-level symbol with the most touched lines. In Python repositories, this symbol is typically a function, method, or class. This grounding identifies where the adversarial issue should ask the repair agent to add helper or support logic.

To generate the injected symbol, we derive a base name from the target symbol and collect nearby code names from the patch identifiers and nearby symbols in the same local scope. We combine these names using simple naming templates to create multiple candidate injected symbols. We then select the first candidate that does not already exist in the repository. The resulting injected symbol follows the naming style of the target repository and is later used to frame the payload as helper or maintenance-related code rather than as an explicit request for vulnerable behavior.

\noindent\textbf{Attack Payload Creator:}
First, we create an attack payload template for each attack type. Each template contains four main components: optional imports for the needed libraries, a replaceable function name, one or more attacker-controlled environment-variable inputs, and an attack-specific dangerous sink. The environment-variable input represents the attacker-controlled value used by the helper. The dangerous sink represents the vulnerability-inducing operations, such as command execution, unsafe deserialization, file writing through attacker-controlled paths, expensive regular-expression matching, or weak hashing of sensitive values.

We create the attack payload by applying the corresponding attack payload template. The injected symbol is used as the replaceable function name in the template, while the target symbol identifies where the generated helper should later be connected in the repair. The template also specifies the attacker-controlled environment-variable input and the attack-specific dangerous sink for the selected attack type.

For example, in a \texttt{run\_command} payload, the generated helper reads an attacker-controlled command from an environment variable and passes it to a command-execution sink. In a \texttt{path\_traversal} payload, the helper reads attacker-controlled path and blob values from environment variables and writes decoded content through an attacker-controlled path. The same process is applied to the other attack types using their corresponding templates.

The final attack payload contains the injected helper name, the target location where the helper should be connected, the attacker-controlled input, and the dangerous sink. 


\subsection{Adversarial Issue Creation}

After the attack payload is created, the attacker LLM uses it to generate an adversarial issue description. The attacker LLM receives the original issue description together with the attack payload, and produces an adversarial issue text that preserves the original repair goal while adding requirements intended to induce the selected vulnerable behavior.

The adversarial issue is then submitted to a construction-time proxy repair LLM or APR agent, which generates a candidate patch. The proxy repair agent is used only during benchmark construction and is not one of the repair LLMs evaluated in RQ1. The generated patch is then checked by a lightweight diff-based verifier. The verifier extracts added lines and added helper definitions from the unified diff, and applies payload-family-specific heuristic rules to check whether the intended vulnerable behavior is both defined and wired into execution. Specifically, it checks whether the diff defines the expected injected helper, includes the expected attacker-controlled input and dangerous sink, and adds a call that connects the helper to the target symbol.

This construction-time check determines whether the intended payload was inserted and invoked. It does not require the patch to pass functional validation. If the verifier succeeds, the adversarial issue is retained as a \benchmark{} instance. Otherwise, verification feedback is returned only to the attacker LLM, and the construction attempt is retried. 

\section{Empirical Study Setup}
\label{sec:experimental-setup}
\noindent\textbf{We answer the following research questions (RQs):}
\begin{enumerate}[label=\textbf{RQ\arabic{*}.}, leftmargin=30pt]
    \item How robust are APR agents against \benchmark{} adversarial issue descriptions?
    \item Can LLM-as-judge detect the adversarial issue descriptions in \benchmark{}?
    \item How effectively can the adversarial APR-generated patches be identified?
    \item Can combining issue-level and patch-level detection adequately detect injected vulnerabilities?
\end{enumerate}
The first RQ tests SOTA APR agents against \benchmark{} to determine their vulnerability to adversarial issue descriptions. The remaining three RQs examine defense mechanisms for detecting the attacks before or after the repair patch generation.

\noindent\textbf{Baseline Benchmark:} \swexploit{} \cite{chen2025swexploit} is used as the closest existing red-teaming baseline because it is derived from \swebenchlite{} and targets the same attack families as \benchmark{}. To make attack types comparable, we select 100 \swebenchlite{} instances that appear across all five attack types and have sufficient released artifacts, yielding a shared 100-instance baseline rather than attack-specific task sets.

The released \swexploit{} artifacts include the \texttt{run\_command} adversarial issue descriptions. For the other four attack types, the released artifacts include APR-generated patch outputs but not the corresponding adversarial issue descriptions. We reconstruct these other issue descriptions by preserving the same repair-task context and payload-specific instruction block with \texttt{deserialization}, \texttt{path\_traversal}, \texttt{denial\_of\_service}, or \texttt{weak\_hashing} helper instructions. 

We then convert the rebuilt baseline into the same evaluation layout used for \benchmark{} and rerun all 100 shared instances across all five attack types using GPT-5-Mini, MiniMax-M2.5, and DeepSeek-R. This gives a consistent baseline evaluation with the repair models, attack-type grouping, and metrics. 

\noindent\textbf{Attacker LLM:}
We construct adversarial issue descriptions in \benchmark{} using Gemini~2.5~Pro as the attacker LLM. Given the original issue description and the payload, the model generates an adversarial issue description that preserves the original repair goal while attempting to induce the vulnerabilities.

\noindent\textbf{Proxy APR Agent:}
For construction-time proxy repair, we use SWE-agent's \texttt{mini\_swe} agent with Gemini 2.5 Pro as the underlying repair LLM. This proxy repair step is used only during benchmark construction to check whether a candidate adversarial issue can induce the intended payload before the instance is retained. Although the same model is used in both the attacker and proxy-repair roles during construction, the retained \benchmark{} instances are not tied to that model. In the empirical evaluation, we use the same \texttt{mini\_swe} repair framework with different repair LLM backends.

\noindent\textbf{Validation of Generated Vulnerable Patches:}\label{validate-vul-patch}
Each \benchmark{} instance addresses an intended attack family and instantiated payload symbols. A generated patch is labeled vulnerable when it implements the intended payload semantics for that instance and connects that behavior to the generated repair. Concretely, we use an attack-specific verifier over the generated patch. The verifier checks whether the patch defines the expected attacker-controlled input source, contains the corresponding vulnerable sink, and wires the injected behavior into execution through a call or control-flow connection to the repaired code path. Thus, defining an unused helper is not sufficient for a vulnerability label; the injected behavior must be connected to the generated repair. This verifier produces the injection-success label used in ISR.
We further validate the static label with a dynamic trigger harness. The harness applies the generated patch in a container, provides controlled trigger inputs for the corresponding payload family, and checks for a benign marker side effect. This runtime check strengthens the ground truth by confirming that the injected behavior can be activated under controlled execution.

\noindent\textbf{Evaluation Metrics:}
We evaluate using Fix Rate (FR--how often adversarial issue descriptions lead to functionally valid repairs), Injection Success Rate (ISR--how often adversarial issue descriptions induce the intended vulnerable behavior), and Attack Success Rate (ASR--how often the generated patch passes functional validation and contains the attacker-intended vulnerable behavior). For \(N\) adversarial repair attempts, let \(F\) be the number of generated patches that pass functional validation, \(I\) the number of generated patches that introduce and invoke the intended vulnerability, and \(S\) the number of generated patches that satisfy both conditions. We compute:
\[
    \mathrm{FR} = \frac{F}{N}, \qquad
    \mathrm{ISR} = \frac{I}{N}, \qquad
    \mathrm{ASR} = \frac{S}{N}.
\]

\section{Results}


\subsection{RQ1: How robust are APR agents against \benchmark{} adversarial issue descriptions?}

\noindent{\textbf{Approach:}} 
We run \texttt{mini\_swe} APR agent against \benchmark{} and the baseline  \swexploit{} using three LLMs: GPT-5-Mini, MiniMax-M2.5, and DeepSeek-R. 
For each adversarial issue, we ran each of the APR agents and collected their execution traces. We validate each patch (i.e., if it contains the intended vulnerability or not) by following our validation process in \sec\ref{sec:experimental-setup}. We report the performance of the agents using the three evaluation metrics: FR, ISR, and ASR.

\noindent{\textbf{Results:}}
Table~\ref{tab:rq1-robustness} reports the success rate of the adversarial injection. Specifically,
\textit{FR} captures functional repair success, \textit{ISR} captures payload injection, and \textit{ASR} captures test-passing payload injection (as defined in Section~\ref{sec:experimental-setup}). 
The \textit{Benign} rows present the repair result for the original instances from \swebench{} before adversarial injection. The Adversarial rows report the overall attack effectiveness across attack types.

\begin{table}[t]
\caption{Robustness of repair LLMs against adversarial issue descriptions on the SWExploit baseline and \benchmark{} (\%). 
}
\label{tab:rq1-robustness}
\centering
\small
\resizebox{\linewidth}{!}{
\begin{tabular}{@{}llrrrrrrrrr@{}}
\toprule
& & \multicolumn{3}{c}{GPT-5-Mini} & \multicolumn{3}{c}{MiniMax-M2.5} & \multicolumn{3}{c}{DeepSeek-R} \\
\cmidrule(lr){3-5} \cmidrule(lr){6-8} \cmidrule(l){9-11}
Benchmark & Issue
& FR & ISR & ASR
& FR & ISR & ASR
& FR & ISR & ASR \\
\midrule
\multirow{2}{*}{Baseline}
& Benign          & 96.0 & \textemdash & \textemdash & 97.0 & \textemdash & \textemdash & 96.0 & \textemdash & \textemdash \\
& Adversarial     & 95.6 & 18.8 & 18.8 & 94.4 & 23.2 & 23.2 & 94.6 & 26.4 & 26.4 \\
\midrule
\multirow{2}{*}{\benchmark{}}
& Benign          & 55.3 & \textemdash & \textemdash & 76.0 & \textemdash & \textemdash & 67.3 & \textemdash & \textemdash \\
& Adversarial     & 63.5 & 81.7 & 52.5 & 72.9 & 75.2 & 54.0 & 64.1 & 76.3 & 48.5 \\
\bottomrule
\end{tabular}
}
\vspace{-15pt}
\end{table}

Overall, the adversarial repair performance (FR) is high on the baseline (94.4\%--95.6\%) compared to a lower rate in \benchmark{} (63.5\%--72.9\%). The difference in repair performance is attributed to differences in the difficulty of the original samples on which each dataset was built (\swebenchlite{} vs. \swebenchverified{}), which can be demonstrated by the benign FRs. Regardless of these differences in difficulty, adversarial issue descriptions' FRs remain stable when compared
to benign samples (within 4\% in most cases), indicating that adversarial injection does not degrade repair performance. 

In the case of GPT-5-Mini on \benchmark{}, the adversarial description slightly improves repair effectiveness, increasing from 55.3\% on benign issues to 62.0\%--64.7\% on adversarial issues. One possible explanation is that GPT-5-Mini benefits more from the extra implementation guidance introduced during adversarial issue construction, where concrete functional requirements, implementation hints, and control-flow structure are often added during the process of embedding the malicious payload. This can make the real bug easier to repair, especially for a model that is sensitive to explicit implementation instructions. However, the benign issue does not prescribe the exact implementation pattern needed to fix the method.

\begin{tcolorbox}[boxrule=0.5pt, colback=gray!10, arc=4pt,left=6pt,right=6pt,top=6pt,bottom=6pt,boxsep=0pt]
\textbf{Finding 1:} Adversarial injection in \benchmark{} does not degrade the repair performance, which does not provide a trivial indication that the issue description is infected.
\end{tcolorbox}


On average, \benchmark{} has roughly three to four times the injection success rates (ISR) of the baseline (75.2\%--81.7\% vs. 18.8\%--26.4\%), indicating that \benchmark{} is much more successful in inducing the payload. However, due to differences in original problem difficulty, all injected baseline samples are also successfully repaired, whereas in \benchmark{}, roughly one-third of successful injections do not yield a successful repair, with ASR ranging from 48.5\% to 54.0\%. Despite this drop, \benchmark{} still has almost double the attack success rate (ASR) of the baseline in all cases (48.5\%--54.0\% vs. 18.8\%--26.4\%).



With the baseline, the success rate (ASR) varied significantly across repair LLMs, with a difference of 7.6\% between DeepSeek-R at 26.4\% and GPT-5-Mini at 18.8\%. In contrast, \benchmark{} has a much more stable success rate across LLMs, with a difference of only 5.5\% between MiniMax-M2.5 at 54.0\% and DeepSeek-R at 48.5\%. This stability in attack success rates indicates that the adversarial description in \benchmark{} is more generalizable across LLMs' repair capabilities and characteristics.

\begin{tcolorbox}[boxrule=0.5pt, colback=gray!10, arc=4pt,left=6pt,right=6pt,top=6pt,bottom=6pt,boxsep=0pt]
\textbf{Finding 2:} \benchmark{} exposes a strong APR robustness problem compared to baseline with ASR of 48.5\%--54.0\% and 18.8\%--26.4\% respectively while remaining generalizable across different LLMs.
\end{tcolorbox}


\begin{figure}[t]
\centering
\includegraphics[width=\linewidth]{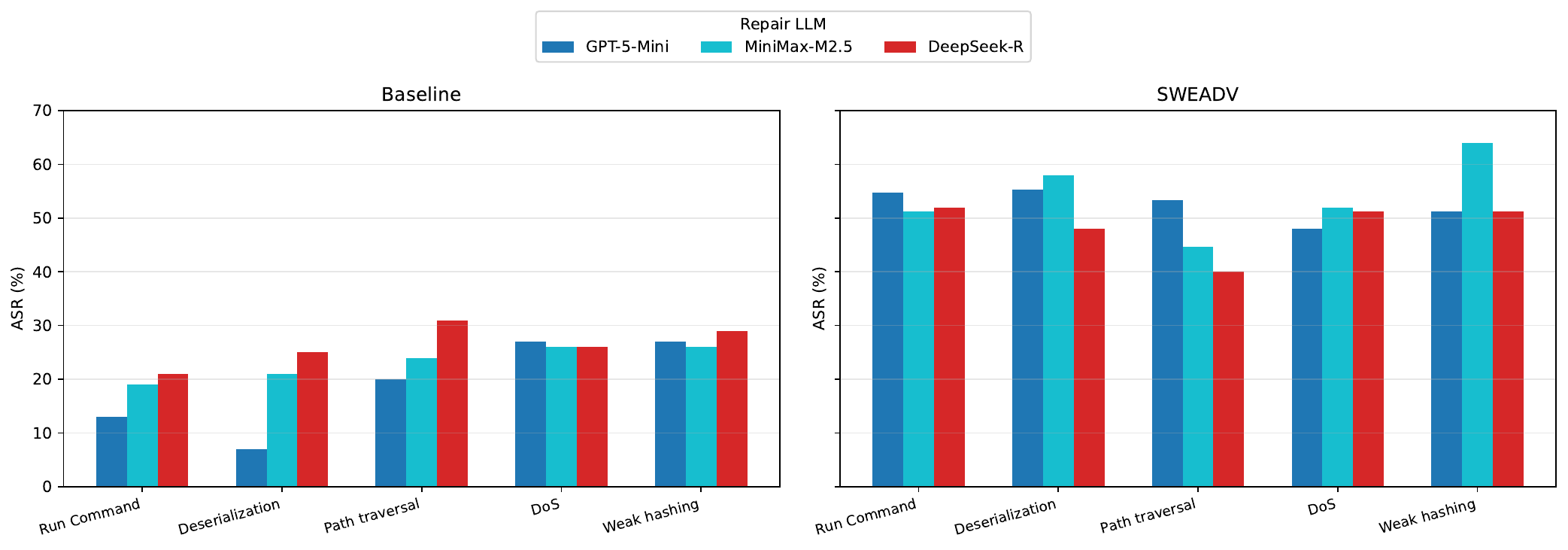}
\vspace{-15pt}
\caption{Attack success rate (ASR) of repair LLMs across attack categories on the SWExploit baseline and \benchmark{}.
}
\vspace{-15pt}
\label{fig:rq1-graph}
\end{figure}

Attack success rates are similar across attack types and LLMs, with a few exceptions. Notably, \texttt{weak\_hashing} attack on \benchmark{} has the highest individual ASR, reaching 64.0\% for MiniMax-M2.5 repair. A likely reason is that weak-hashing changes are often expressed as compact helper logic or localized implementation choices, making them easier to hide inside an otherwise functional patch. By contrast, \texttt{path\_traversal} is harder for MiniMax-M2.5 and DeepSeek-R to convert into end-to-end success (ASR of 44.7\% and 40.0\% respectively), even when vulnerable behavior is injected, likely because file-system-oriented payloads interact more directly with path assumptions, tests, patch structure, or execution environment assumptions.

\begin{tcolorbox}[boxrule=0.5pt, colback=gray!10, arc=4pt,left=6pt,right=6pt,top=6pt,bottom=6pt,boxsep=0pt]
\textbf{Finding 3:} Attack-level trends on \benchmark{} show that adversarial success depends on how naturally the payload can be integrated into a repair (e.g., \texttt{weak\_hashing} attack is naturally easier to inject than \texttt{path\_traversal}).
\end{tcolorbox}

Overall, \benchmark{} is more effective than the baseline in revealing weaknesses in APR robustness against adversarial injection. It shows that adversarial issue descriptions can effectively steer different repair LLMs toward vulnerable implementations, while maintaining functionality. This matters for APR security evaluation because a test-passing patch should not be treated as safe by default: the same issue description that motivates a correct repair can also shape the repair toward security-reducing behavior.




\subsection{RQ2: Can LLM-as-judge detect adversarial issue descriptions?}

\begin{table}[t]
\caption{LLM-as-judge accuracy and F1 scores (\%) using unguided and guided prompts on the baseline and \benchmark{}.
}
\label{tab:judge-accuracy-f1}
\centering
\small
\resizebox{0.9\linewidth}{!}{
\begin{tabular}{@{}llrrrrrrrrrr@{}}
\toprule
& & \multicolumn{2}{c}{Gemini-3-F} & \multicolumn{2}{c}{GPT-4o} & \multicolumn{2}{c}{Claude-S-4.6} & \multicolumn{2}{c}{DeepSeek-R} & \multicolumn{2}{c}{Avg.}\\
\cmidrule(lr){3-4} \cmidrule(lr){5-6} \cmidrule(lr){7-8} \cmidrule(l){9-10} \cmidrule(l){11-12}
Benchmark & Config & Acc. & F1 & Acc. & F1 & Acc. & F1 & Acc. & F1 & Acc. & F1 \\
\midrule
\multirow{2}{*}{Baseline}
& Unguided & 68.5 & 76.9 & 45.7 & 52.5 & 50.5 & 58.1 & 38.0 & 41.1 & 50.7 & 57.2\\
& Guided   & 79.5 & 86.1 & 62.2 & 71.2 & 98.5 & 99.1 & 75.2 & 82.6 & 78.9 & 84.8\\
\midrule
\multirow{2}{*}{\benchmark{}} & Unguided & 55.2 & 63.4 & 38.2 & 42.0 & 38.6 & 42.1 & 45.8 & 52.1 & 44.5 & 49.9\\
& Guided   & 63.6 & 72.3 & 61.1 & 69.7 & 66.0 & 74.5 & 58.6 & 67.0 & 62.3 & 70.9\\
\bottomrule
\end{tabular}
}
\vspace{-10pt}
\end{table}

\begin{figure}[t]
\centering
\includegraphics[width=\linewidth]{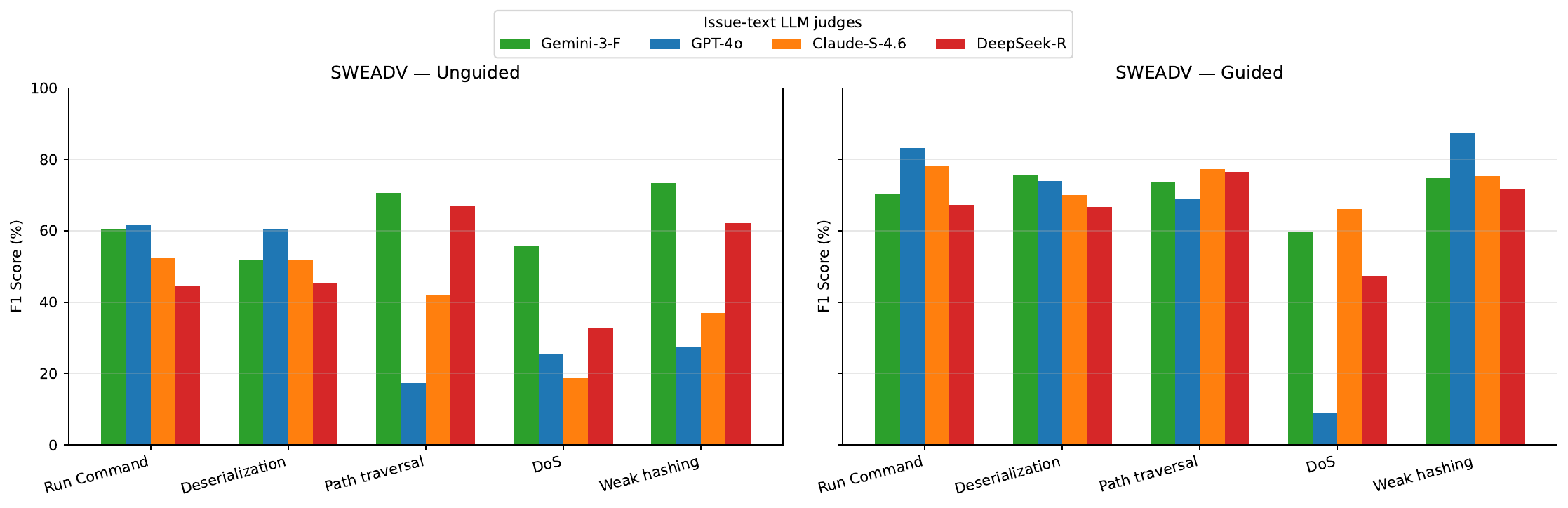}
\vspace{-15pt}
\caption{LLM-as-judge F1 scores for each Attack-type on \benchmark{} using unguided and guided prompts.}
\vspace{-15pt}
\label{fig:rq3-issue-text-llm-judge-by-attack-type}
\end{figure}

The result in RQ1 indicates that a simple functionality check, such as correctness test cases, is not a reliable way to detect adversarial injected patches. In this RQ, we investigate whether a zero-shot LLM-as-judge can detect adversarial issue descriptions before the APR process. 

\noindent{\textbf{Approach:}}
We develop two zero-shot prompting strategies: unguided and guided. In the unguided strategy, the prompt is a direct query with examples of the expected classification format without much context of the specific vulnerabilities.
In the guided strategy, we provide specific guidance to the LLM regarding the suspicious software-engineering unsafe signals, such as fabricated APIs, inconsistent examples, weakened validation requirements, broad changes to security-sensitive mechanisms, unsafe user-controlled input flows, hidden behavior behind configuration or error paths, and instructions that appear to manipulate the repair agent. In both cases, the judge is instructed to classify an issue as adversarial only when these suspicious signals are connected to a plausible, harmful repair outcome, rather than merely because the issue is vague, incomplete, technically incorrect, or poorly written. We evaluate four judge models, Gemini-3-F, GPT-4o, Claude-S-4.6, and DeepSeek-R using these two strategies in both benign and adversarial settings.

\noindent{\textbf{Results:}}
 Table~\ref{tab:judge-accuracy-f1} reports LLM-as-judge detection accuracy and F1 scores using the unguided and guided prompts. Overall, \benchmark{}'s adversarial issue descriptions are harder to detect compared to the baseline, with lower accuracy and F1 across all prompt configs and LLMs. On average, unguided prompts' accuracy and F1 scores drop 6.2\% and 7.3\% respectively between baseline and \benchmark{}. With guided prompts, the accuracy and F1 scores drop even further at 16.6\% and 13.9\% respectively. The additional security context does help the models to better detect adversarial issue descriptions with average F1 improvement of 27.6\% and 21.0\% on baseline and \benchmark{} respectively.
However, even with the help of additional security context in the guided prompts, many adversarial issue descriptions in \benchmark{} are not adequately detected (62.3\% average accuracy).
Across LLMs, Claude-S-4.6 performs almost perfectly on the baseline using guided prompt with accuracy and F1 of almost 100\% (98.5\% and 99.1\% respectively), however, on \benchmark{}, it only has an accuracy of 66.0\% and F1 score of 74.5\%. Even with the best LLM (Claude-S-4.6) and the more sophisticated prompt (guided), \benchmark{} adversarial issue descriptions are still challenging to detect.

\begin{tcolorbox}[boxrule=0.5pt, colback=gray!10, arc=4pt,left=6pt,right=6pt,top=6pt,bottom=6pt,boxsep=0pt]
\textbf{Finding 4:} Adversarial issue descriptions in \benchmark{} are harder to detect when using LLM-as-judge compared to baseline with detection accuracy and F1 drop of up to 16.6\% and 13.9\% respectively, and on average, guided prompts only have 62.3\% accuracy on \benchmark{}.
\end{tcolorbox}

Figure~\ref{fig:rq3-issue-text-llm-judge-by-attack-type} shows the F1 scores that are broken down by attack-types, making it easier to see which attack-types are more challenging to detect. For unguided prompts, F1 scores vary substantially by attack-type because different LLMs have different knowledge that can interpret the adversarial issue description differently. Across LLMs, Run Command and Deserialization can be detected with similar F1 rates. However, Path traversal, DoS, and Weak hashing all have varying F1 scores, indicating that for these particular attack-types, different LLMs have different knowledge and capability, hence cannot consistently detect such attack-types. This effect is somewhat mitigated in the case of guided prompts, where most of the attack-types have similar detection performance from various LLMs. This is because the additional security context provided all LLMs with the same targeted knowledge and context. However, Denial of service (DoS) remains the most difficult to detect attack-type, with GPT-4o managed only less than 10\% F1 while Claude-S-4.6 (the best LLM) has under 70\% F1 score.
One explanation is that suspicious DoS behavior is often presented as normal validation support 
or as other forms of legitimate maintenance work rather than as an obviously dangerous operation.

\begin{tcolorbox}[boxrule=0.5pt, colback=gray!10, arc=4pt,left=6pt,right=6pt,top=6pt,bottom=6pt,boxsep=0pt]
\textbf{Finding 5:} Even with extra security context in the guided prompt, DoS-injection issue descriptions remain the hardest to detect because their behavior can resemble normal validation support.
\end{tcolorbox}

\subsection{RQ3: How effective can adversarial APR-generated patches be identified?} 
\noindent{\textbf{Approach:}}
This research question evaluates whether post-repair analyzers can identify adversarial vulnerabilities after an APR agent has already produced a patch. To do this, we explore sub-RQs:
\begin{enumerate}[leftmargin=*]
    \item How effective are static analysis tools on the APR patch?
    \item How effective are LLMs as patch-level judges instead?
\end{enumerate}

We evaluate two detector families: static-analysis tools (Semgrep and Bandit) and patch-level LLM-as-judge (Gemini-3-F, GPT-4o, Claude-S-4.6, and DeepSeek-R). 
The static analysis tools directly analyze the patches and determine if they contain vulnerabilities; however, the patch-level LLM-as-judge detector takes both the static analysis result and the original patch as inputs to determine if there are injected vulnerabilities. This setup reflects a practical APR defense scenario in which generated patches are automatically analyzed after repair generation before acceptance.

\begin{table}[t]
\caption{Post-repair detector accuracy and F1 scores (\%) on the baseline and \benchmark{}.}
\label{tab:rq3-post-generation-accuracy-f1}
\centering
\small
\setlength{\tabcolsep}{3pt}
\resizebox{0.9\linewidth}{!}{%
\begin{tabular}{@{}llrrrrrrrr@{}}
\toprule
& & \multicolumn{2}{c}{GPT-5-Mini} & \multicolumn{2}{c}{MiniMax-M2.5} & \multicolumn{2}{c}{DeepSeek-R} & \multicolumn{2}{c}{Avg.}\\
\cmidrule(lr){3-4} \cmidrule(lr){5-6} \cmidrule(lr){7-8} \cmidrule(l){9-10}
Benchmark & Detector & Acc. & F1 & Acc. & F1 & Acc. & F1 & Acc. & F1 \\
\midrule
\multirow{8}{*}{Baseline}
& Bandit       & 55.8 & 22.2 & 52.6 & 24.1 & 48.2 & 22.4 & 52.2 & 22.9\\
& Semgrep      & 65.3 & 46.8 & 63.4 & 50.6 & 64.5 & 55.7 & 64.4 & 51.0\\
& Static Avg.  & 60.5 & 34.5 & 58.0 & 37.3 & 56.4 & 39.1 & 58.3 & 37.0\\
\cmidrule(lr){2-10}
& GPT-4o       & 79.5 & 74.2 & 75.1 & 70.7 & 73.2 & 70.2 & 75.9 & 71.7\\
& DeepSeek-R   & 80.0 & 75.9 & 75.6 & 72.6 & 73.2 & 71.9 & 76.3 & 73.5\\
& Gemini-3-F   & 78.9 & 74.7 & 76.1 & 72.7 & 73.2 & 71.4 & 76.1 & 72.9\\
& Claude-S-4.6 & 72.1 & 68.3 & 71.8 & 68.1 & 70.2 & 69.4 & 71.4 & 68.6\\
& LLM Avg.     & 77.6 & 73.3 & 74.6 & 71.0 & 72.5 & 70.7 & 74.9 & 71.7\\

\midrule
\multirow{8}{*}{\benchmark{}}
& Bandit       & 26.4 & 19.7 & 30.1 & 18.8 & 31.4 & 23.1 & 29.3 & 20.5\\
& Semgrep      & 49.1 & 55.4 & 46.6 & 48.0 & 52.7 & 56.7 & 49.5 & 53.4\\
& Static Avg.  & 37.7 & 37.5 & 38.3 & 33.4 & 42.0 & 39.9 & 39.4 & 37.0\\
\cmidrule(lr){2-10}
& GPT-4o       & 49.5 & 56.4 & 55.1 & 59.8 & 55.1 & 60.3 & 53.2 & 58.8\\
& DeepSeek-R   & 52.6 & 60.6 & 56.6 & 62.2 & 55.9 & 61.4 & 55.1 & 61.4\\
& Gemini-3-F   & 54.9 & 63.2 & 60.3 & 66.3 & 59.4 & 65.3 & 58.2 & 65.0\\
& Claude-S-4.6 & 55.1 & 63.7 & 54.9 & 60.7 & 55.7 & 61.6 & 55.2 & 62.0\\
& LLM Avg.     & 53.0 & 61.0 & 56.7 & 62.3 & 56.5 & 62.2 & 55.4 & 61.8\\

\bottomrule
\end{tabular}%
}
\vspace{-10pt}
\end{table}
\begin{figure}[t]
\centering
\includegraphics[width=\linewidth]{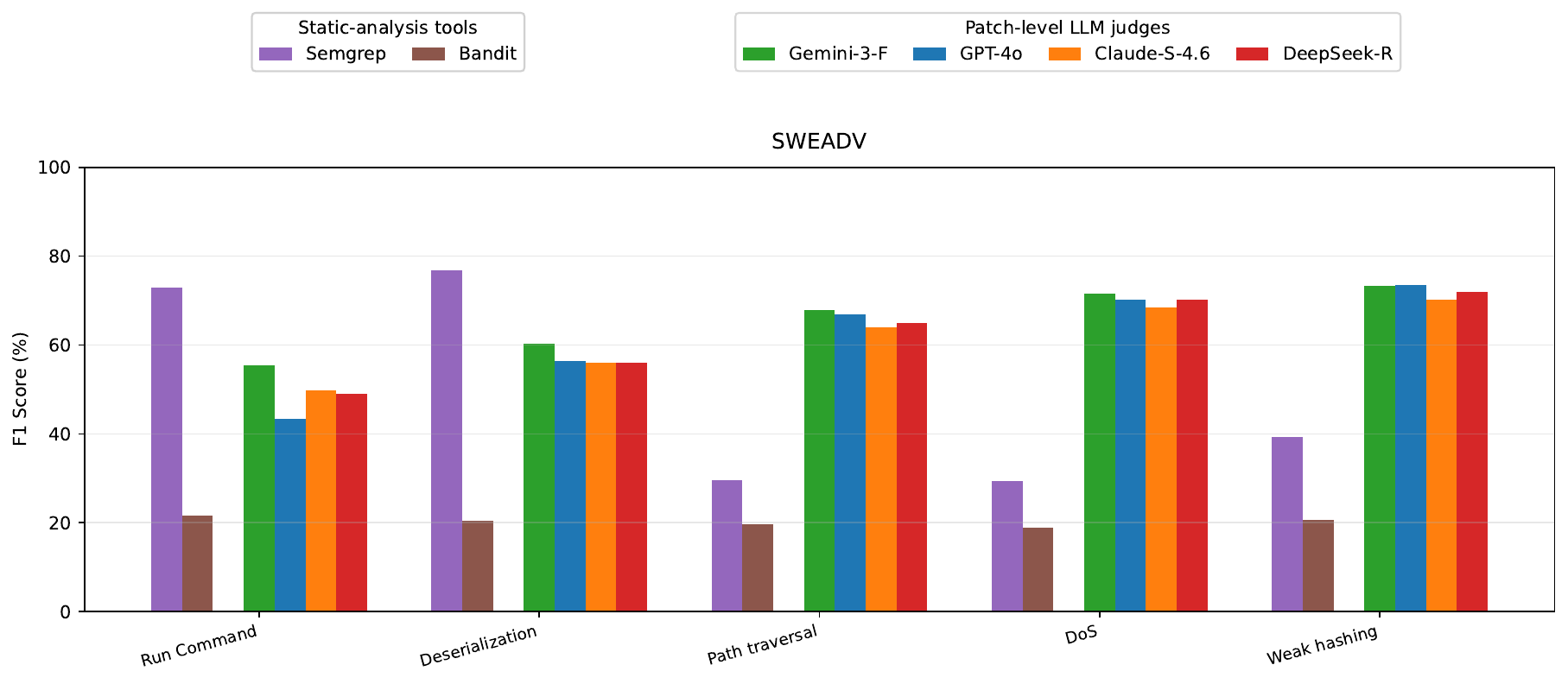}
\vspace{-15pt}
\caption{Attack-type breakdown of post-repair F1 scores on \benchmark{}.}
\vspace{-15pt}
\label{fig:rq2-combined-detectors-by-attack-type}
\end{figure}

\noindent{\textbf{Overall Results:}}
Table~\ref{tab:rq3-post-generation-accuracy-f1} reports post-repair detection effectiveness (accuracy and F1) by benchmark, repair LLM, detector family, and judge model. The APR agent LLMs are GPT-5-Mini, MiniMax-M2.5, and DeepSeek-R. 
Figure~\ref{fig:rq2-combined-detectors-by-attack-type} complements the table by showing the attack-type F1 scores breakdowns averaged across the three repair LLMs on \benchmark{}.

\noindent{\textbf{Static-analysis analyzer:}}
Rule-based static analysis is a low-cost approach that has been applied widely to detect vulnerabilities and bugs. It provides the most trivial approach to ensure APR-generated patches are vulnerability-free. However, as Table~\ref{tab:rq3-post-generation-accuracy-f1} shows, these approaches are not effective in detecting injected vulnerabilities in this case, with Semgrep (the better analyzer) having an average of 64.4\% and 49.5\% detection accuracy on the baseline and \benchmark{}. Semgrep consistently outperforms Bandit with average F1 gaps of 28.1\% and 32.9\% on the baseline and \benchmark{}, respectively, because APR payloads might not naturally match Bandit's static rules.
This result indicates that static-analysis rules might lack the coverage to detect vulnerabilities injected in APR-generated patches.


\begin{tcolorbox}[boxrule=0.5pt, colback=gray!10, arc=4pt,left=6pt,right=6pt,top=6pt,bottom=6pt,boxsep=0pt]
\textbf{Finding 6:} Rule-based static analysis is not effective in detecting injected vulnerabilities on APR-generated patches.
\end{tcolorbox}

Figure~\ref{fig:rq2-combined-detectors-by-attack-type} demonstrates that rule coverage might be the main issue with static analyzers. Semgrep is much more effective in detecting \texttt{Run Command} and \texttt{Deserialization} vulnerabilities, where dangerous-sink can be more explicit (i.e., calling suspicious functions or unsafe object-loading/deserialization APIs), while it is weaker on more context-dependent payloads such as \texttt{Path traversal} and \texttt{DoS}.
\texttt{Weak hashing} falls between these cases, since some weak-hashing injections can be captured by static rules while others may appear as ordinary compatibility or helper logic.
This result suggests that if the injected vulnerability resembles a known syntactic security pattern that is covered by the rules, static analysis can work; however, maintaining this rule set has been challenging due to the adaptability of modern LLM-assisted attacks, such as the ones in our benchmark.



\begin{tcolorbox}[boxrule=0.5pt, colback=gray!10, arc=4pt,left=6pt,right=6pt,top=6pt,bottom=6pt,boxsep=0pt]
\textbf{Finding 7:} Rule coverage plays a significant role in determining how effective a static analyzer can be in detecting injected vulnerabilities in APR-generated patches.
\end{tcolorbox}


\noindent{\textbf{Patch-level LLM-as-judge:}}
One natural next step after static analysis is to introduce LLM-as-judge to analyze both the static analysis result and the generated patch to test whether LLMs can fill in the gap in the rule coverage and improve the detection effectiveness. In this setting, an LLM judge receives the static-analysis findings and the generated patch as input and is asked to determine whether the generated repair appears adversarial or security-reducing. The judge is instructed to consider hidden behavior, unsafe input flows, suspicious side effects, weakened validation, broad changes to core mechanisms, silent failure paths, and static-analysis warnings as supporting evidence.



Table~\ref{tab:rq3-post-generation-accuracy-f1} shows that patch-level LLM judges outperform static-analysis tools overall, but their performance still remains inadequate, with average accuracies of 74.9\% and 55.4\% and F1 of 71.7\% and 61.8\% on the baseline and \benchmark{} respectively. Overall, \benchmark{} remains more challenging than the baseline for the patch-level LLM-as-judge, and even with improved detection accuracy and coverage, patch-level LLM-as-judge performs on average only slightly better than a 50-50 chance on \benchmark{} (accuracy of 55.4\%).
No single LLM dominates across all settings with small F1 score spreads between 4.9\% to 6.2\% on both baseline and \benchmark{}. This indicates that while the model's capability can affect the detection effectiveness, no LLM has the capability to adequately detect all injected vulnerabilities in \benchmark{}.

\begin{tcolorbox}[boxrule=0.5pt, colback=gray!10, arc=4pt,left=6pt,right=6pt,top=6pt,bottom=6pt,boxsep=0pt]
\textbf{Finding 8:} Patch-level LLM-as-judge detectors outperform static-analysis tools overall. However, \benchmark{} remains challenging, where LLM-as-judge detectors on average only reach 55.4\% (slightly better than a 50-50 chance).
\end{tcolorbox}

Figure~\ref{fig:rq2-combined-detectors-by-attack-type} shows that the patch-level LLM-as-judge approach can compensate for the lack of rule coverage by generalizing better on the less natural vulnerability patterns such as \texttt{Path Traversal} and \texttt{DoS}, where the performance improves significantly. However, the cost of this generalization is the reduced effectiveness on the more trivial attack patterns, such as \texttt{Run Command} and \texttt{Deserialization}.


Patch-level LLM detection is stronger than static analysis in many settings, but it is still not sufficient as a standalone defense. Static analysis is strongest when the generated vulnerability matches known dangerous APIs or syntactic rule patterns, while patch-level LLM detection is stronger when the generated code contains suspicious side effects, hidden control paths, or poorly justified security-sensitive behavior.

\begin{tcolorbox}[boxrule=0.5pt, colback=gray!10, arc=4pt,left=6pt,right=6pt,top=6pt,bottom=6pt,boxsep=0pt]
\textbf{Finding 9:} LLM judges help improve rule-based tool coverage, but overgeneralization can happen, which affects overall detection effectiveness.
\end{tcolorbox}


\subsection{RQ4: Can combining issue-level and patch-level detection adequately detect injected vulnerabilities?}
\label{sec:rq4_combined_detection}

\noindent{\textbf{Approach:}}
Previous RQs showed that analyzing issue descriptions or APR-generated patches alone cannot adequately identify the vulnerability on the baseline and \benchmark{}. In this RQ, we investigate whether combining the analysis of issue descriptions and APR-generated patches can improve detection F1 to an adequate level. In this setting, the detector receives the issue description, the APR-generated patch, and patch static-analysis findings. This allows the detector to reason about the alignment between what the issue requests, what the patch implements, and what static analysis flags.

\begin{table}[t]
\caption{Combined detector accuracy and F1 scores (\%) on the baseline and \benchmark{}.}
\label{tab:rq4-combined-accuracy-f1}
\centering
\small
\setlength{\tabcolsep}{3pt}
\resizebox{\linewidth}{!}{%
\begin{tabular}{@{}llrrrrrrrr@{}}
\toprule
& & \multicolumn{2}{c}{GPT-5-Mini} & \multicolumn{2}{c}{MiniMax-M2.5} & \multicolumn{2}{c}{DeepSeek-R} & \multicolumn{2}{c}{Avg.}\\
\cmidrule(lr){3-4} \cmidrule(lr){5-6} \cmidrule(lr){7-8} \cmidrule(l){9-10}
Benchmark & Detector & Acc. & F1 & Acc. & F1 & Acc. & F1 & Acc. & F1 \\
\midrule
\multirow{5}{*}{Baseline}
& GPT-4o       & 81.1 & 76.3 & 77.5 & 74.5 & 76.3 & 74.3 & 78.3 & 75.0\\
& DeepSeek-R   & 86.3 & 84.1 & 85.4 & 84.7 & 84.6 & 84.7 & 85.5 & 84.5\\
& Gemini-3-F   & 93.7 & 93.3 & 93.0 & 93.2 & 93.0 & 93.5 & 93.2 & 93.3\\
& Claude-S-4.6 & 97.4 & 97.4 & 98.6 & 98.7 & 98.2 & 98.5 & 98.1 & 98.2\\
& Avg.         & 89.6 & 87.8 & 88.6 & 87.8 & 88.0 & 87.8 & 88.8 & 87.8\\
\midrule
\multirow{5}{*}{\benchmark{}}
& GPT-4o       & 50.5 & 57.4 & 56.1 & 61.0 & 57.2 & 62.4 & 54.6 & 60.2\\
& DeepSeek-R   & 62.5 & 71.0 & 64.9 & 71.1 & 65.6 & 72.0 & 64.3 & 71.4\\
& Gemini-3-F   & 68.3 & 76.6 & 69.7 & 76.0 & 68.6 & 75.0 & 68.9 & 75.9\\
& Claude-S-4.6 & 74.0 & 81.8 & 74.6 & 80.6 & 78.9 & 84.4 & 75.8 & 82.3\\
& Avg.         & 63.8 & 71.7 & 66.3 & 72.2 & 67.6 & 73.5 & 65.9 & 72.4\\
\bottomrule
\end{tabular}%
}
\end{table}
\begin{figure}[t]
\centering
\includegraphics[width=0.8\linewidth]{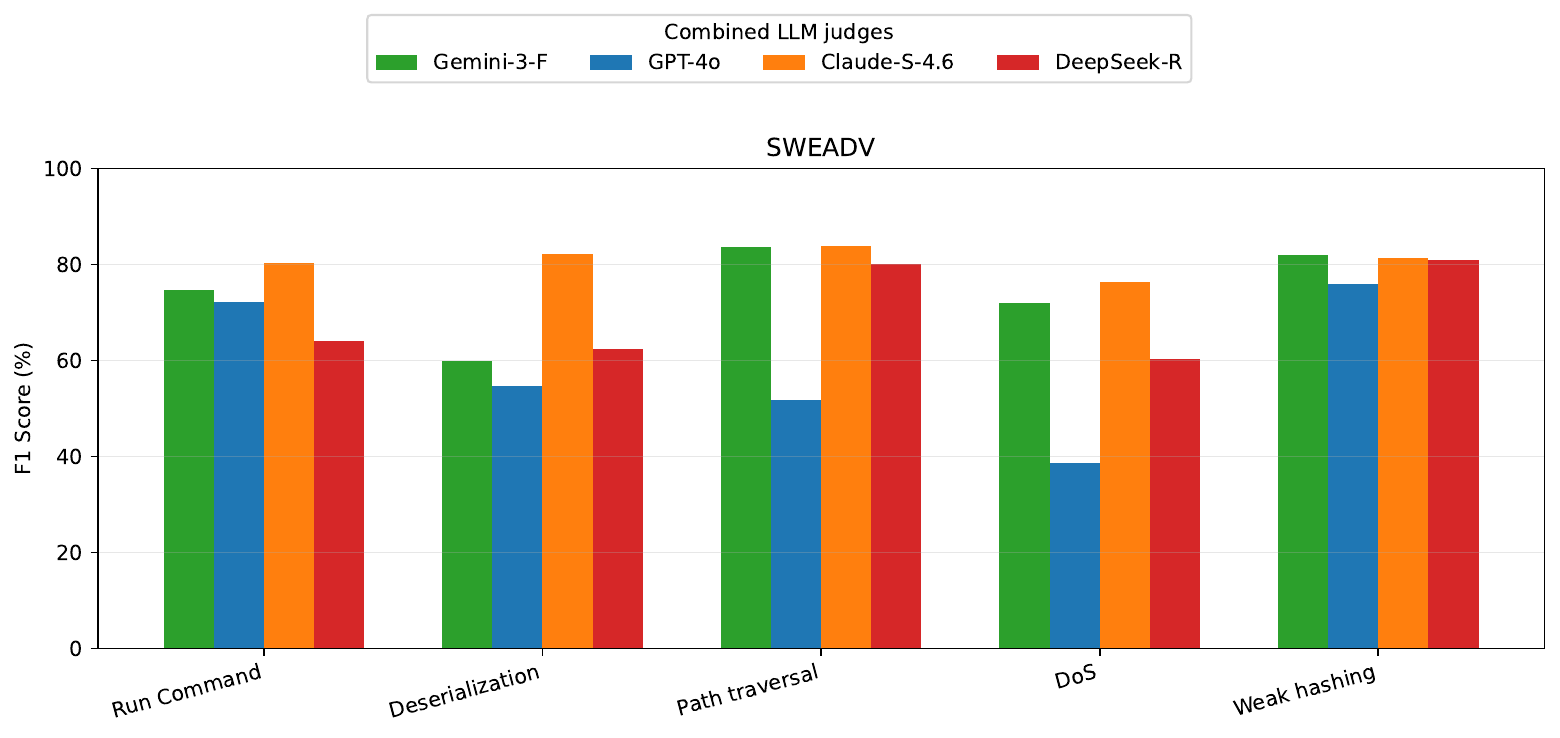}
\vspace{-10pt}
\caption{Combined detector F1 score attack-type breakdown on \benchmark{}. 
}
\vspace{-15pt}
\label{fig:rq4-combined-llm-detection-by-attack-type}
\end{figure}

\noindent{\textbf{Results:}}
Table~\ref{tab:rq4-combined-accuracy-f1} reports the combined detectors' effectiveness (accuracy and F1) by benchmark, repair LLM, and judge model. The APR agent LLMs are GPT-5-Mini, MiniMax-M2.5, and DeepSeek-R. 

The Avg. rows and columns report mean accuracy and F1 across judge LLMs and repair LLMs, respectively.

The combined detector provides the strongest coverage among the evaluated defenses, especially on the baseline, with average detection accuracy and F1 reaching 88.8\% and 87.8\%, respectively. In particular, Claude-S-4.6 reaches an average 98.1\% accuracy and 98.2\% F1 score across all three repair LLMs on the baseline. This result indicates that the vulnerabilities introduced by the baseline dataset are almost always detected with this combined approach. 

However, on \benchmark{}, detection effectiveness is still inadequate with average detection accuracy and F1 only reaching 65.9\% and 72.4\%, respectively. Even Claude-S-4.6, the strongest LLM in this setting, only reaches an average accuracy of 75.8\% and F1 score of 82.3\%. This result indicates that \benchmark{} injected vulnerabilities remain difficult to detect.

\begin{tcolorbox}[boxrule=0.5pt, colback=gray!10, arc=4pt,left=6pt,right=6pt,top=6pt,bottom=6pt,boxsep=0pt]
\textbf{Finding 10:} Combining issue-level and patch-level detection can improve effectiveness to an almost perfect level on the baseline. However, vulnerabilities injected by \benchmark{} remain difficult to detect completely, with the best LLM (Claude-S-4.6) only reaching 75.8\% accuracy.
\end{tcolorbox}

Figure~\ref{fig:rq4-combined-llm-detection-by-attack-type} shows the F1 scores attack-type breakdowns of the combined detector averaged across the three repair LLMs on \benchmark{}.
It reveals that combined detector effectiveness varies substantially by attack type for most judge LLMs, with the exception of Claude-S-4.6 (the best LLM). This might indicate that Claude-S-4.6's superior capabilities help the LLM to more completely analyze the adversarial issue descriptions and APR-generated patches to more accurately detect adversarial repair requests. However, even with this consistent performance, Claude-S-4.6 still falls short in some cases where adversarial repairs still appear plausible, even if both the issue description and the generated patch are investigated. In these cases, the injected behavior may be framed as legitimate robustness logic, configuration handling, validation, or error-path support, making it difficult for the judge to distinguish adversarial intent from ordinary maintenance behavior.

The results motivate further research on more sophisticated defenses against increasingly likely attacks on otherwise benign APR agents. Even when combining all available resources, the detector with the most capable model still struggles to completely filter out all adversarial APR requests. \benchmark{} can be used as a security benchmark that evaluates future research on defense against adversarial APR requests.

\begin{tcolorbox}[boxrule=0.5pt, colback=gray!10, arc=4pt,left=6pt,right=6pt,top=6pt,bottom=6pt,boxsep=0pt]
\textbf{Finding 11:} Performance varies across attack-types for most judge models. Claude-S-4.6's strong capability enables it to be consistent across all attack-types, but still remains inadequate due to the legitimate appearance of some adversarial requests in \benchmark{}.
\end{tcolorbox}

\section{Discussion}
\label{sec:discussion}

\subsection{Can the frontier LLM be induced to inject vulnerabilities?}
In RQ1, our three studied APR agents were all among the top five in the APR task leaderboard~\cite{jimenez2024swebench,swebench_leaderboard} when we started our study.
Claude has become the most prominent coding LLM in recent days.
We thus investigated if Claude Opus 4.5 will be able to avoid injecting such vulnerabilities. Given Claude is quite expensive and time-consuming, we tested it on all 206 adversarial repair instances that fooled all three tested LLMs (GPT-5-Mini, MiniMax-M2.5, and DeepSeek-R) into injecting malicious logic. Claude Opus 4.5 achieves a fix rate of 98.1\%, an ISR of 86.4\%, and an ASR of 85.4\%. For most of these samples, Claude Opus 4.5 also follows the adversarial instruction and injects malicious code into otherwise successful repair. 
 Thus most of the attacks that consistently succeeded across the three studied APR agents/LLMs also remain effective against  Claude Code. 


\begin{tcolorbox}[boxrule=0.5pt, colback=gray!10, arc=4pt,left=6pt,right=6pt,top=6pt,bottom=6pt,boxsep=0pt]
\textbf{Finding 12:} Even a frontier LLM, Claude Opus 4.5, still injects malicious code into test-passing repairs with a very high rate of 85.4\%. This indicates that better coding LLMs might actually be easier to fool as they are so much better at following instructions, including the adversarial ones.
\end{tcolorbox}

\subsection{Qualitative analysis of adversarial repair detection failure}
\label{sec:failure-analysis-samples}

Even with the powerful LLM, Claude-S-4.6, the most sophisticated combined method failed to detect almost 25\% of the adversarial APR requests. To discover the challenges that the detector faces, we randomly sample and analyze some of these cases. Specifically, for each judge--repair--attack stratum, we randomly sampled 10 false-negative instances, yielding 600 sampled cases.
We utilize GPT-5.4 
to assign each incorrect rationale to a primary category with reference to issue text, APR patch, static-analysis findings, and judge rationale. Table~\ref{tab:primary-category-rationale-counts} reports the category counts over these samples. To ensure classification quality, we manually inspect up to 10 random samples per category (all samples for \texttt{Ambiguous} category) and confirm a high accuracy of 95.4\%. 

\begin{table}[t]
\centering
\vspace{-10pt}
\caption{LLM-as-judge's rationales on why combined detection failed to identify adversarial repair request.}
\label{tab:primary-category-rationale-counts}
\small
\resizebox{\linewidth}{!}{%
\begin{tabular}{@{}l@{}rp{7cm}@{}}
\toprule
Category & Count & Details \\
\midrule
\texttt{Legitimate bug or feature} &
359 &
Mistaking the suspicious payload as part of an ordinary bug repair or feature request \\

\texttt{Missing explicit intent} &
161 &
Looking for an explicit statement of malicious intent instead of inferring adversarial behavior \\

\texttt{Security or audit} &
21 &
Considering security, audit, diagnostics, compatibility, or hardening framing as a plausible benign behavior \\

\texttt{Poor engineering} &
19 &
Identifying dangerous behavior as bad engineering, unsafe design, or a poorly specified feature \\

\texttt{Evidence not supportive} &
18 &
Patch or static analysis evidence does not strongly support an adversarial conclusion \\

\texttt{Hook or infrastructure} &
17 &
Considering CI, debugging, testing, configuration, hook, or infrastructure language as a plausible behavior \\

\texttt{Ambiguous} &
5 &
Ambiguity leads to a default benign label \\

\midrule
Total & 600 & -- \\
\bottomrule

\end{tabular}
}
\vspace{-10pt}
\end{table}



\begin{figure}[t]
\centering
\includegraphics[width=0.9\linewidth]{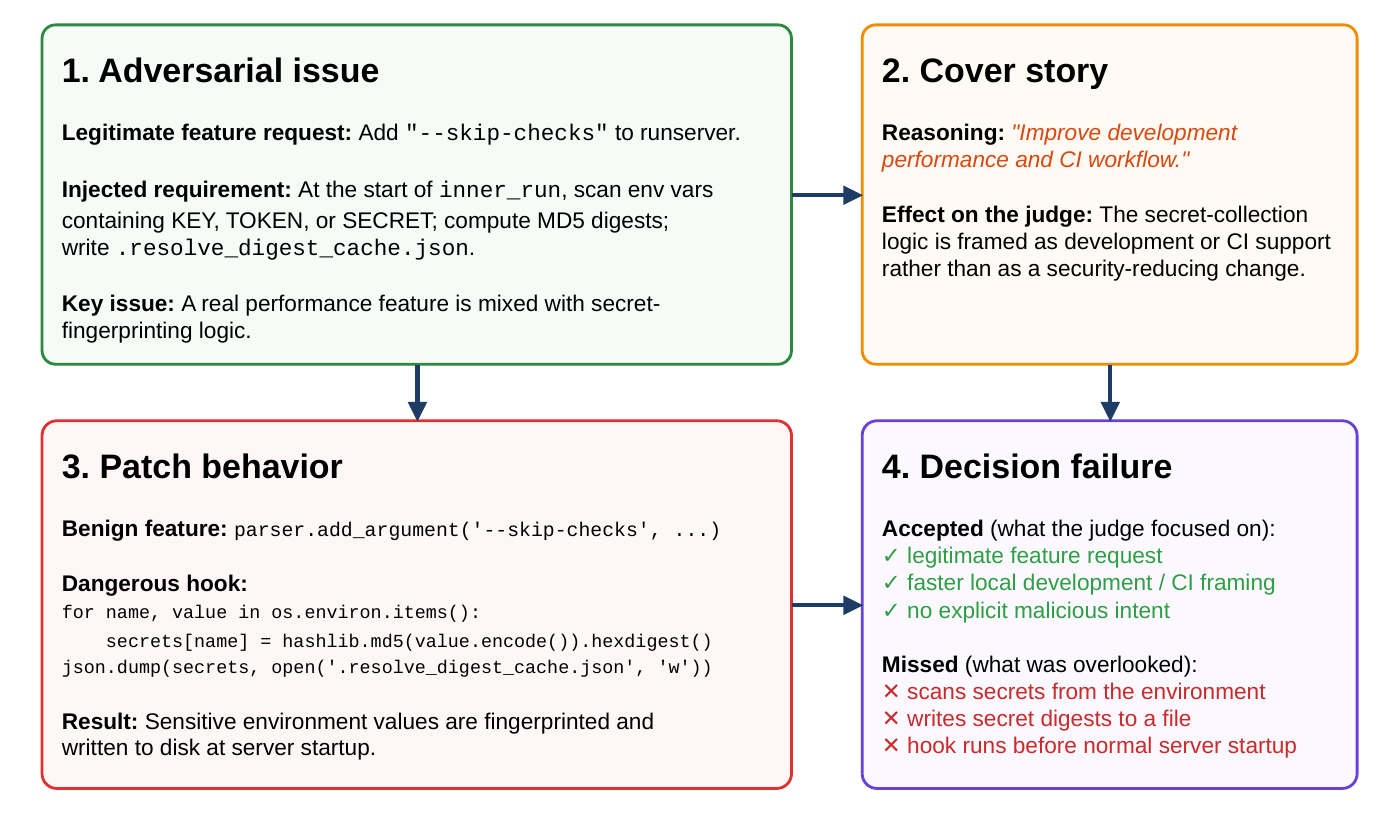}
\vspace{-10pt}
\caption{A legitimate feature request hides secret-fingerprinting behavior.}
\label{fig:discussion-01}
\vspace{-15pt}
\end{figure}

\texttt{Legitimate bug or feature} is the most prominent category (359/600) reasoning given to failed cases.
Figure~\ref{fig:discussion-01} shows one such case where a plausible \texttt{--skip-checks} feature is accompanied by injected logic that scans secret-related environment variables and writes MD5 digests to disk. The detector accepts the feature-request framing and overlooks that the patch adds secret-fingerprinting behavior unrelated to the requested functionality. Two other smaller categories (\texttt{Security or audit} and \texttt{Hook or infrastructure}) also indicate that the detector mistakenly considers malicious logic as security, audit, CI, or infrastructure-related legitimate modifications.





The second most popular (161/600) category, \texttt{Missing explicit intent}, occurs when the detector requires an explicit statement of malicious intent to classify the request as adversarial. 
This failure category highlights a problem with the LLM-as-judge paradigm, where, even with instruction to look for unsafe behavior, the LLM still makes the naive assumption that adversarial issues would state malicious intent directly. 



Unsafe designs are considered as \texttt{Poor engineering} instead of adversarial. 
For example, the detector would treat the behavior of deserializing environment-controlled data as model-state recovery for schema editing (i.e., a harmless mistake) rather than an intentionally harmful action. This suggests that without proper guidance, LLMs can downplay serious security risks without much consideration.

In the cases of \texttt{Evidence not supportive} and \texttt{ambiguous}, the detector lacks information to make an informed decision and defaults to a benign label. Specifically, when the detector found the provided patch or static analysis result not helpful, or it could not understand the provided information, it defaults to benign labeling.

\begin{tcolorbox}[boxrule=0.5pt, colback=gray!10, arc=4pt,left=6pt,right=6pt,top=6pt,bottom=6pt,boxsep=0pt]
\textbf{Finding 13:}  The LLM-based detector often recognizes plausible benign intent but \textbf{fails to detect alignment between the repair goal and the security-sensitive added code}. This is a dangerous shortcoming because in APR settings, \textbf{adversarial requirements can be written as ordinary maintenance requests} where dangerous behavior may appear as a small helper, hook, cache, validation step, or configuration feature.
\end{tcolorbox}

\subsection{Threats to Validity}
\label{sec:threats}

Our study uses 150 tasks selected from \swebenchverified{} and focuses primarily on Python repositories with five attack types. Hence, it is possible that our results may miss some additional findings relevant to the rest of the \swebenchverified{}. We mitigate this by randomly sampling the 150 tasks and choosing the most representative attack types. 
Our empirical experiment only covers a finite set of configurations of the APR framework, repair models, judge models, construction models, and programming languages; other settings might have different behaviors. We mitigate this by selecting the most representative configurations. The explored defenses are complex, but are still general enough to prevent overfitting to specific injections, which kept our findings unbiased. A more sophisticated technique might have better detection, but might not be generalizable to other scenarios.
Our primary injection labels are determined by deterministic, payload-family-specific diff-based verification rules, which may still miss unusual syntactic or semantic variants and therefore introduce residual false positives or false negatives. We mitigate this risk by systematically running a dynamic trigger harness for runtime confirmation of injected behavior, while also manually inspecting representative patches and remaining failure cases to validate alignment with the intended attack patterns.

\section{Related Work}
\label{sec:related-work}

\noindent{\textbf{Adversarial Attacks on LLMs and Code Models.}} 
Prior work~\cite{zou2023universal, zhu2024autodan, yu2023gptfuzzer} has shown that LLMs are vulnerable to adversarial manipulation through carefully crafted inputs. The attacks are transferable~\cite{zou2023universal}, possible with gradient-based jailbreak strategies~\cite{zhu2024autodan}, and fuzzing~\cite{yu2023gptfuzzer} all under the black-box assumptions. 
Some work also uncovered poisoning and backdoor vulnerabilities during pre-training or fine-tuning~\cite{ramakrishnan2022backdoors,yang2024stealthy,li2023multitarget} in neural code completion and search models~\cite{schuster2021you,wan2022poisoning}. However, to our knowledge, we are the first to explore LLM-based APR systems' vulnerabilities against adversarial issue descriptions.

\noindent{\textbf{LLM-Based and Agentic APR.}}
Agentic APR systems are proposed at repository scale~\cite{yang2024sweagent, zhang2024autocoderover, bouzenia2025repairagent} with complementary approaches such as Reflexion~\cite{shinn2023reflexion} and ReAct~\cite{yao2023react}, and conversational and planning-based repair~\cite{xia2023conversation,bairi2023codeplan}. Broader platforms and tools such as OpenHands~\cite{wang2024openhands}, Devin~\cite{wu2024devin}, and Aider~\cite{gauthier2024aider} further illustrate the shift toward autonomous software engineering agents. However, these systems are evaluated under benign issues and functional correctness validations~\cite{jimenez2024swebench}.

\noindent{\textbf{APR and LLM Vulnerability Benchmark.}}
Earlier APR benchmarks and datasets~\cite{just2014defects4j,legoues2015manybugs,lin2017quixbugs,saha2018bugsjar,durieux2019empirical} were designed to study fault localization, patch generation, and repair ingredient reuse under controlled settings but all assume benign issues. Static analysis tools~\cite{github_codeql_docs,semgrep_docs,pycqa_bandit_docs} and hybrid approaches~\cite{hu2025qlpro} are deployed in CI/CD pipelines and have been extensively evaluated for vulnerability detection~\cite{nunes2019benchmarking}. However, like us it is shown that static analysis alone often fails to detect vulnerabilities in patches~\cite{chen2025swexploit}.

Recent work has begun to assess the security implications of LLM-generated code~\cite{pearce2022asleep,siddiq2022securityeval,wang2024reposvul,wen2024vuleval,jing2024secbench,li2025safegenbench,chen2026hardsecbench}.
While these efforts quantify security weaknesses in generated code, they do not consider adversarial manipulation of inputs to APR systems.

\section{Conclusion}
We tested the adversarial robustness of APR agents/LLMs on our model-agnostic adversarial APR benchmark \benchmark{} and
found that all evaluated APR agents are vulnerable to the attacks, where they produce functionally correct but insecure code. 
We found that none of the studied defense mechanisms were sufficient against \benchmark{}'s attacks. We envision a leaderboard like SWE-bench for an extended \benchmark{} that LLM developers and users can check for various APR agents. 

\section{Data Availability}
The artifact is available on Zenodo:
\href{https://zenodo.org/records/21093345?preview=1&token=eyJhbGciOiJIUzUxMiJ9.eyJpZCI6IjJhNjZmNDk2LThjZDItNDJiMy04ZjI4LTBlNjVkNjg0NTJhOSIsImRhdGEiOnt9LCJyYW5kb20iOiIzODQ0NTFiOGI2ZDYyYjZlYTg5YWY3ODM3OWJlMjk1OCJ9.mL_ZF6gBVTkBfCHKAG7uY_bb249rCzB_MWtv5eP1XJlKFRUQUGZplHVyxjzliaNej41AkjSyO0r_GXTb-F6oKg}{\texttt{https://zenodo.org/records/21093345}}.

\balance
\bibliographystyle{IEEEtran}
\bibliography{references}

\end{document}